\documentclass[a4paper,11pt]{article}
\usepackage{graphicx}
\usepackage[dvipsnames]{xcolor}
\usepackage[colorlinks,linkcolor=blue,anchorcolor=blue,citecolor=blue]{hyperref}
\usepackage{amsmath}
\usepackage{amssymb} 
\usepackage{geometry} 

\usepackage[backend=biber, style=nature, doi=false]{biblatex} 
\begin{document}

\begin{center}
  
\textbf{\huge{Generation of S-shaped photonic hooks from microcylinders with engineered surface patches}}

\vspace{0.5cm}

\noindent {\large Chu Xu$^{1,2}$, Fen Tang$^{1,2}$, Qingqing Shang$^{1,2}$, Yao Fan$^{2}$, Jiaji Li$^{2}$, Songlin Yang$^{3}$, Dong Wang$^{1}$, Sorin Melinte$^{4}$, Chao Zuo$^{2}$, Zengbo Wang$^{5,6}$, Ran Ye$^{1,2}$*}

\end{center}

\noindent $^{1}$ School of Computer and Electronic Information, Nanjing Normal University, Nanjing 210023, China\\
\noindent $^{2}$ Smart Computational Imaging Laboratory (SCILab), School of Electronic and Optical Engineering, Nanjing University of Science and Technology, Nanjing 210094, China\\
\noindent $^{3}$ Advanced Photonics Center, Southeast University, Nanjing 210096, China\\
\noindent $^{4}$ Institute of Information and Communication Technologies, Electronics and Applied Mathematics, Universit\'{e} Catholique de Louvain, 1348 Louvain-la-Neuve, Belgium\\
\noindent $^{5}$ School of Computer Science and Electronic Engineering, Bangor University, Bangor LL57 1UT, UK \\
\noindent $^{6}$ z.wang@bangor.ac.uk\\
\noindent \textit{*ran.ye@njnu.edu.cn}

\paragraph{ABSTRACT:} Photonic hooks (PHs) are non-evanescent light beams with a highly concentrated curved optical fields. Since their discovery, PHs always have one single inflection point and thus have a hook-like structure. In this work, a new type of PHs with two inflection points and S-shaped structures (S-PHs) were reported for the first time. We theoretically studied the effects of various physical parameters on the generation of S-PHs. Furthermore, we showed that decorating particles with multiple patches can significantly enhance the curvature and length of the S-PHs. The S-PHs may have potential applications in super-resolution imaging, sub-wavelength micromachining, particle and cell manipulation, etc.

\section{Introduction}

Photonic nanojets (PJs) are high-intensity light waves with a subwavelength beam waist generated at the shadow side of illuminated dielectric particles~\cite{04ChenZ}, which have promising applications in various fields such as micromaching~\cite{20YanB},  super-resolution imaging~\cite{11WangZ, 13YeR, 16WangF, 16YangH}, etc. In recent years, a great effort has been put into particle designing to generate PJs with new features~\cite{17LukYanchukBS, 20MininIV3}. In 2015, Minin et al. theoretically discovered a new type of PJs with curved optical fields~\cite{15MininI}. They called them photonic hooks (PHs). PHs was later experimentally proved at terahertz frequency~\cite{19MininIV}, and then at optical frequency~\cite{20MininIV}. PHs have potential applications in various fields, such as nanoparticle and cell manipulation~\cite{19DholakiaK, 20MininIV3} and super-resolution microscopy~\cite{21ShangQ}. In 2021, Shang et al. reported a contrast-enhanced super-resolution imaging technique using patchy microspheres~\cite{21ShangQ}. The patchy microspheres convert incident light into a PH and induce a near-field asymmetric illumination condition. Asymmetric illumination is a common technique in computational microscopic imaging, which can enhance the contrast of objects~\cite{21FanY}. Minin et al. also reported a PH-based terahertz microscopy technique~\cite{21MininOV}.

When light waves pass through a dielectric particle whose shape or refractive index distribution is asymmetrical to the propagation direction of light, the difference in phase velocities and the interference of light waves lead to a curved high-intensity focus, i.e., a PH~\cite{18YueL}. Illuminating geometrically asymmetric dielectric particles, such as dielectric trapezoids~\cite{18YueL, 19MininIV} and glass cubes~\cite{18YangJ} embedded dielectric cylinders, with plane waves is a common way to generate PHs. Particles with an anisotropic refractive index distribution, such as Janus particles~\cite{19GuG, 20GeintsYE}, are also suitable for the generation of PHs. In addition, PHs can be generated by partial or nonuniform illumination of symmetric and homogeneous particles~\cite{20MininIV, 20LiuCY2}, or by using dielectric particles partially covered with opaque films~\cite{21TangF}. For example, Tang et al. used 1 - 35 $\mu$m diameter patchy microcylinders to generate PHs. Half of the cylinder surface is covered with silver films, and PHs with a bending angle of $\sim$ 28.4$^\circ$ can be generated under plane wave illumination~\cite{21TangF}.

However, the aforementioned methods can only generate PHs with one inflection point. In this work, we show that a new type of PHs with two inflection points, S-shaped PHs (S-PHs), can be generated using patchy particles. To the best of our knowledge, it is the first time that S-shaped PHs are reported. Numerical simulations based on the finite-difference-time-domain (FDTD) method were performed to investigate the characteristics of the S-PHs. By changing the background refractive index, particle diameters and the position and coverage ratio of patches, the curvature, maximum intensity and the hot spot position of S-PHs can be effectively tuned. Moreover, the curvature of S-PHs can be further significantly enhanced using patchy particles with multiple patches.

\section{Results and discussion}

\begin{figure}[h!]
\centering\includegraphics[width=\textwidth]{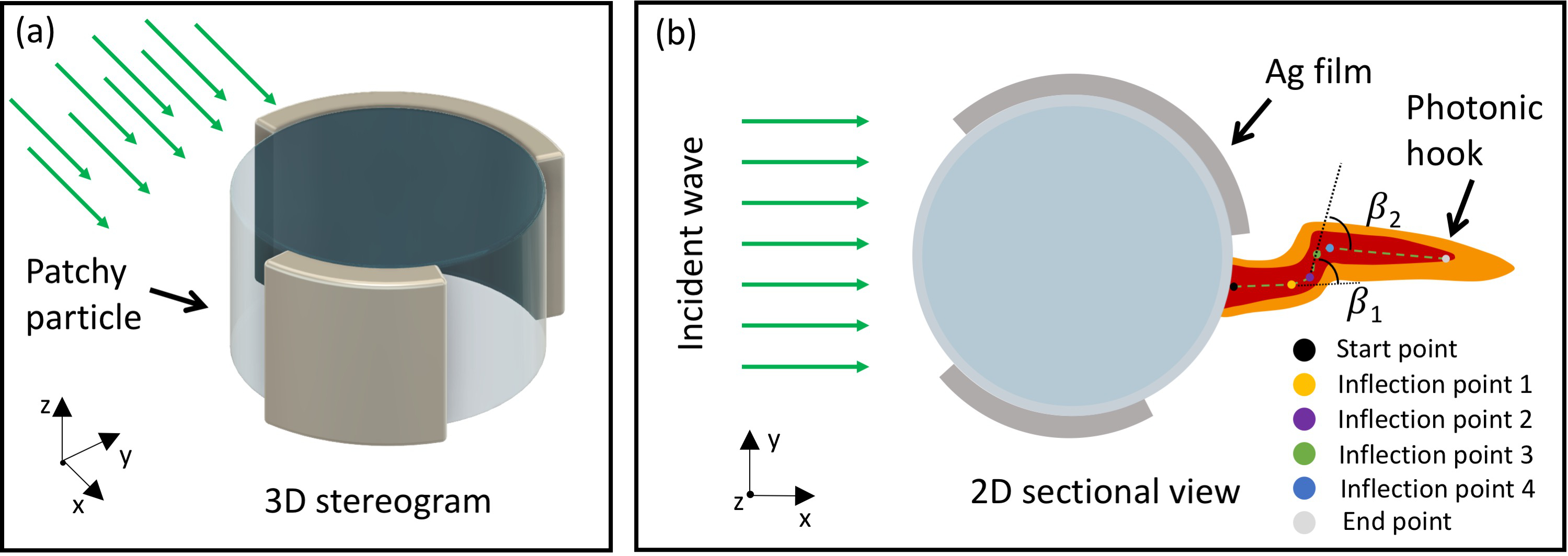}
\caption{Schematics of a patchy microcylinder illuminated by plane waves: (a) 3D stereogram and (b) 2D sectional view.}
 \label{fig:shiyitu}
\end{figure}

Figures~\ref{fig:shiyitu}(a), (b) are the schematic drawings of the 3D stereogram and 2D sectional view of the investigated model. A dielectric microcylinder was created for two-dimensional simulation with the FDTD method using Lumerical FDTD solutions. The surface of the cylinder is partially covered with 100 nm-thick Ag films. The center of the microcylinder is always located at the X = 0 $\mu$m, Y = 0 $\mu$m, and the profiles of the microcylinder as well as the scale bars are given in the following images. As shown in Figure~\ref{fig:shiyitu}(b), an intense focusing of light will occur on the shadow side of the cylinder when a P-polarized monochromatic plane wave ($\lambda$ = 550 nm) propagating parallelly to the X axis passes through the cylinder. In this study, the microcylinder has a constant RI of 1.90, the same as the refractive index of BaTiO$_3$, which is a high-index dielectric material widely used in microsphere-assisted super-resolution imaging~\cite{16WangF, 20YanB, 21ShangQ}. The diameter of the cylinder changes between 1 - 5 $\mu$m and the RI of the background changes between 1.00 - 1.52. For the entire computational domain, non-uniform meshes with RI-dependent element size were used and all of them are smaller than $\lambda$/30.

As shown in Figure~\ref{fig:shiyitu}(b), a typical S-PH starts out with an approximately smooth midline. The midline bends upwards when meeting the S-shaped curve and then becomes stable. After a certain distance, the midline bends downwards to leave the S-shaped curve and then becomes smooth again until reaching the end point of the S-PH. The structure of a S-shaped photonic hook can be approximately defined by the start point (SP), the four inflection points (IP$_1$, IP$_2$, IP$_3$ and IP$_4$) and the end point (EP) of the photonic hook. The inflection points are the points at which the curvature state of the midline changes~\cite{21GuG}. The end point in this study is defined as the point on the middle line of the photonic hook with an intensity enhancement factor of I$_{max}$/e ~\cite{20MininIV, 20LiuCY}. The I$_{max}$ is the largest ${\mid}E{\mid}^2$ enhancement formed at the shadow side of the patchy microcylinder. Based on these points, the curvature of the S-shaped photonic hooks can be defined by the bending angle $\beta_1$ and $\beta_2$. $\beta_1$ is the angle between the two lines respectively connecting the SP with the IP$_1$, and the IP$_2$ with the IP$_3$. $\beta_2$ is the angle between the two lines respectively connecting the IP$_2$ with the IP$_3$, and the IP$_4$ with the EP.

\section{Results and Discussion}

\begin{figure}[htb]
    \centering
    \includegraphics[width=\textwidth]{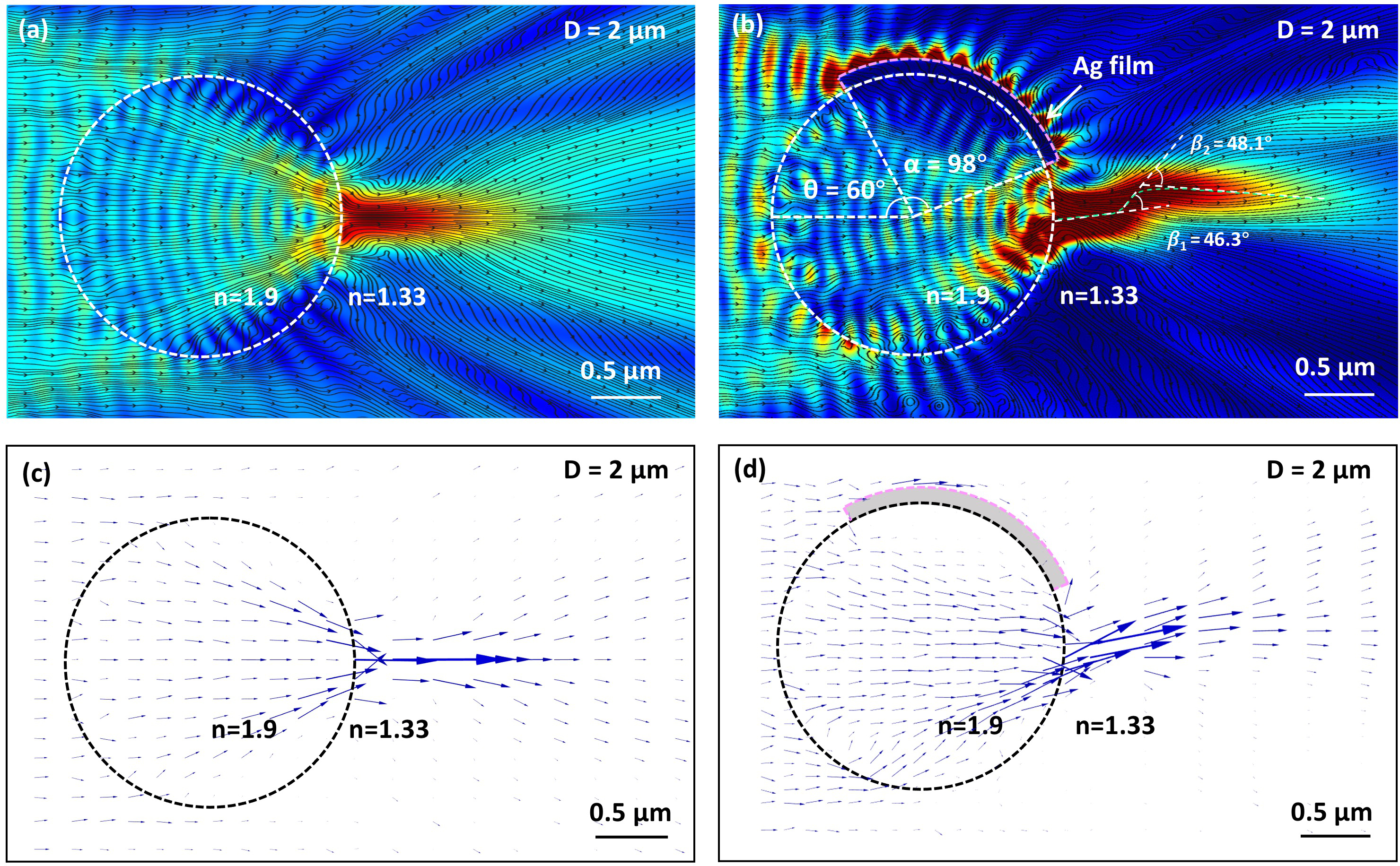}
    \caption{(a) PNJ formed by a 2 $\mu$m-diameter microcylinder; (b) S-PH formed by a 2 $\mu$m-diameter patchy microcylinder; (c,d) Poynting vectors of (c) the PNJ and (d) the S-PH.} 
    \label{fig:pnjvsph}
\end{figure}

First, we compared the optical fields of 2 $\mu$m-diameter pristine and patchy microcylinders under plane wave illumination. The background medium has a RI of n=1.33. As shown in Figure~\ref{fig:pnjvsph}(a), the incident light passing through the pristine cylinder forms a conventional PNJ on the shadow side of the cylinder. The generated PNJ has a symmetric $\mid$E$\mid^2$ distribution with the midline of the PNJ as the center of symmetry. On the contrary, as for the patchy microcylinder shown in Figure~\ref{fig:pnjvsph}(b), the upper part of the incident light is blocked by the Ag film on the top surface of the cylinder. The rotation and opening angle of the film is $\theta$=60$^\circ$ and $\alpha$=98$^\circ$, respectively. We find that a curved light beam with a S-shaped structure, i.e., a S-PH, with bending angles of $\beta_1$=46.3$^\circ$ and $\beta_2$=48.1$^\circ$ is formed at the shadow side of the cylinder.

As reported in the previous work~\cite{19GuG, 20LiuCY}, the formation mechanism of PNJs and PHs can be analyzed using the time-averaged Poynting vector. In this work, the Poynting vector (blue conical arrows) of the optical field formed by pristine and patchy cylinders under plane wave illumination is simulated with the FDTD method [Figures~\ref{fig:pnjvsph}(c), (d)], and the corresponding field-lines of the Poynting vector distribution are shown as the black lines in Figures~\ref{fig:pnjvsph}(a) and (b). As shown in Figure~\ref{fig:pnjvsph}(c), the spatial distribution of the Poynting vector inside and near the pristine cylinder is symmetric to the midline of the PNJ [Figure~\ref{fig:pnjvsph}(c)]. Because the length of the conical arrows is proportional to the value of energy flux, the area containing longer arrows indicates a higher energy flux in that area. We can see that the energy flow corresponding to the pristine cylinder's optical field is focused into a classical PNJ at the shadow side of the cylinder [Figure~\ref{fig:pnjvsph}(c)]. However, as for patchy cylinders, part of the incident light is reflected backwards to the space by the Ag film, which breaks the symmetry of illumination and makes the energy flow inside the microcylinder unbalanced. This asymmetric flow of energy is then focused into a curved beam after leaving the patchy cylinder, as shown in Figures~\ref{fig:pnjvsph}(b),(d).

\begin{figure}[htb]
    \centering
    \includegraphics[width=\textwidth]{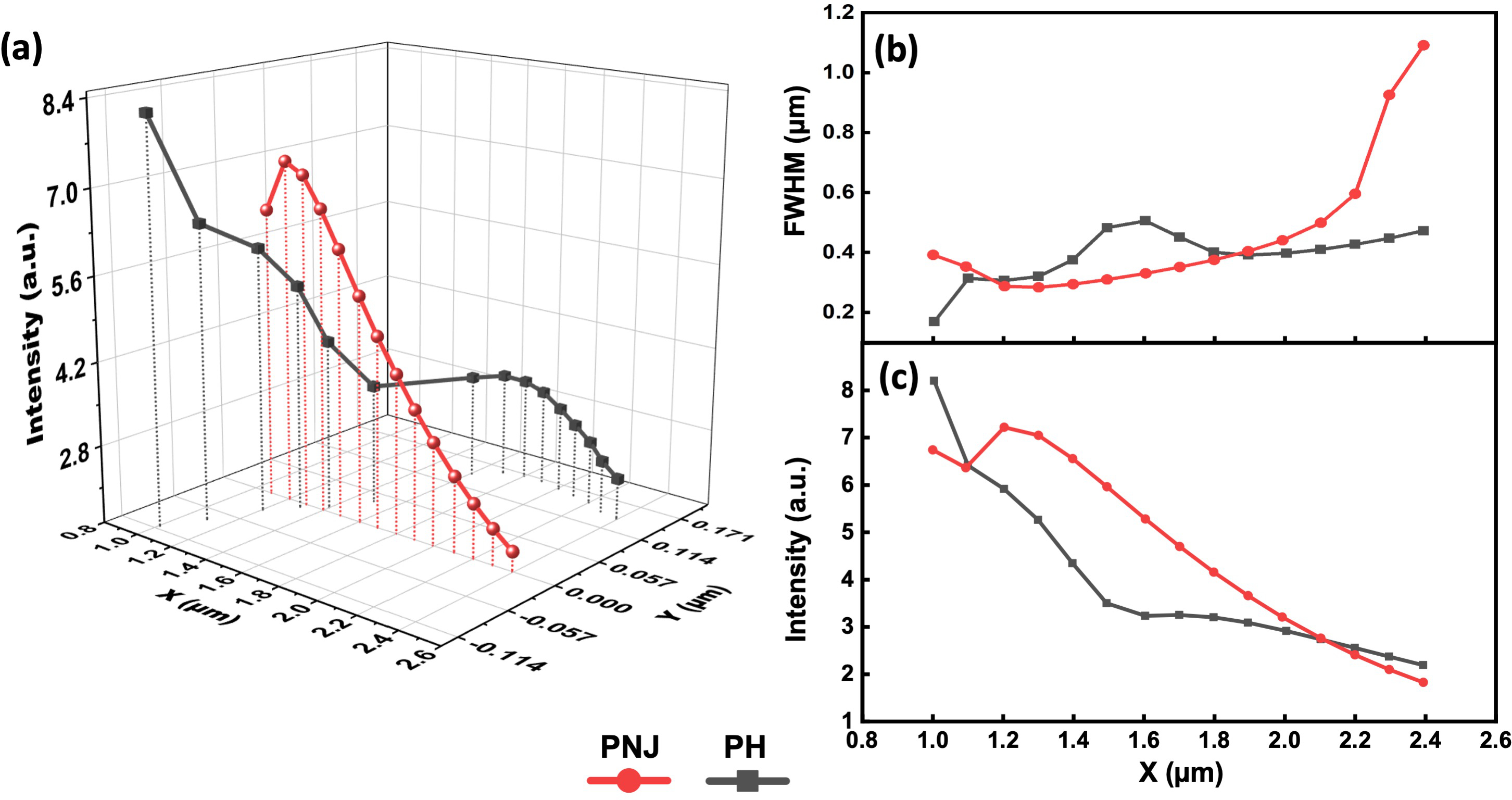}
    \caption{(a) 3D trajectory of I$_{max}$ points of a PNJ and a S-PH.  Intensity profiles of a PNJ and S-PH formed by a 2 $\mu$m-diameter microcylinder; (b) FWHM of the PNJ and the S-PH at different X positions; (c) Maximum intensity of the PNJ and S-PH at different X positions.} 
    \label{fig:pnjvsph2}
\end{figure}

Figure~\ref{fig:pnjvsph2}(a) is the 3D trajectory of the I$_{max}$ points of the PNJ [Figure~\ref{fig:pnjvsph}(a)] and the S-PH [Figure~\ref{fig:pnjvsph}(b)] obtained by slicing along X-axis in the range of X = 1 $\mu$m to X = 2.4 $\mu$m. Due to the symmetric nature of the PNJ, the optical field of the PNJ has the strongest intensity at its center at Y = 0 $\mu$m. The I$_{max}$ point of the photonic hook, on the other hand, has a S-shaped profile and a shift in the Y-axis. Figure~\ref{fig:pnjvsph2}(b) shows the FWHM of the intensity profiles of the PNJ and S-PH at different x-coordinates. The FWHM at the edge of the microcylinders (X = 1.0 $\mu$m) is 0.17 $\mu$m (0.31 $\lambda$) and 0.39 $\mu$m (0.71 $\lambda$) for the S-PH and PNJ, respectively. After the light propagating another 0.2 $\mu$m along the X-axis, the FWHM of the PNJ decreases to 0.29 $\mu$m (0.53 $\lambda$) and the light reaches the focal point of the cylinder. The S-PH has a slightly larger FWHM (0.56 $\lambda$) at X = 1.2 $\mu$m, and its beam waist continues to increase with the propagation distance of light. After a propagation distance of 1 $\mu$m, the S-PH shows a significantly narrower beam waist than the PNJ, and the FWMH of the S-PH  (0.85 $\lambda$) is $\sim$ 43\% of the PNJ (1.98 $\lambda$) at X = 2.4 $\mu$m. The property that the S-PH can retain a subwavelength FWHM at a longer propagation distance could be beneficial to the applications of particle manipulation and microfabrication. Figure~\ref{fig:pnjvsph2}(c) shows the variation of peak intensity of the PNJ (red line) and the S-PH (black line) along the X-axis. The S-PH shows a better intensity enhancement performance as the I$_{max}$ of the S-PH and PNJ is 8.20 and 7.22, respectively. The S-PH also have a higher intensity at propagation distance longer than 4 $\lambda$. The peak intensity obtained at X = 2.4 $\mu$m is 2.19 for the S-PH and 1.83 for the PNJ, respectively.

\begin{figure}[htb]
    \centering
    \includegraphics[width=\textwidth]{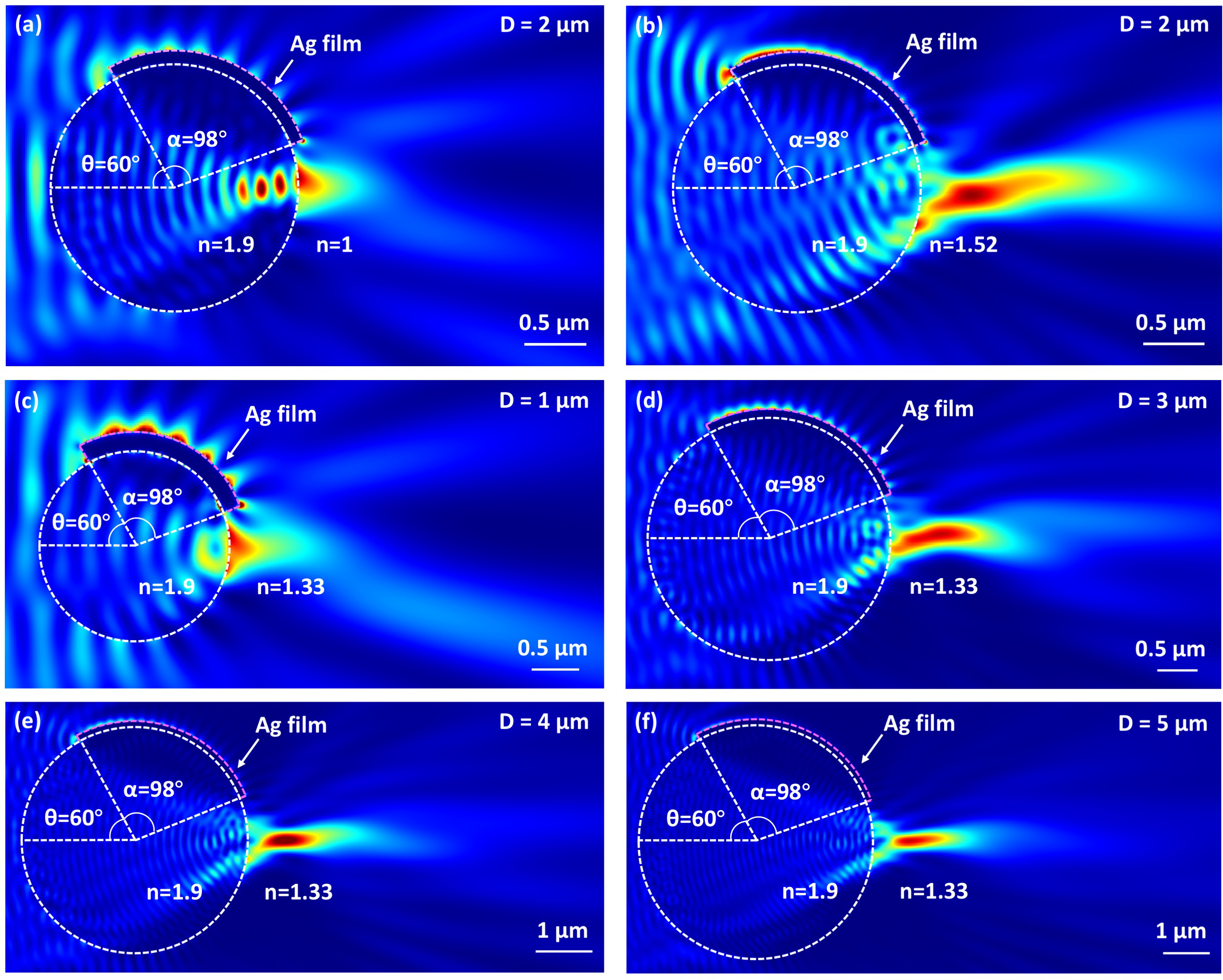}
    \caption{(a), (b) Optical fields generated by 2 $\mu$m-diameter patchy microcylinders at a background RI of (a) n = 1.00, (b) n = 1.52; (c)-(f) Optical fields generated by patchy microcylindes with a diameter of (c) D = 1 $\mu$m, (d) D = 3 $\mu$m, (e) D = 4 $\mu$m, (f) D = 5 $\mu$m.} 
    \label{fig:bjdx}
\end{figure}

As shown in Figures~\ref{fig:bjdx}(a),(b), the influence of background RI on the characteristics of obtained PHs was studied by changing the background RI to n=1.00 [Figure~\ref{fig:bjdx}(a)] and n=1.52 [Figure~\ref{fig:bjdx}(b)]. The particle diameter was kept constant at 2 $\mu$m. The rotation angle $\theta$ and the opening angle $\alpha$ of the Ag films are 60$^\circ$ and 98$^\circ$, respectively. As shown in Figure~\ref{fig:bjdx}(a), when the background RI is 1.0, the light entering the microcylinder is focused to the edge of the cylinder. A higher RI contrast between the particle and the background leads to a stronger bending ability. The I$_{max}$ is at X = 1.00 $\mu$m, Y = 0.09 $\mu$m with a value of 5.78. The intensity enhancement decreases to I$_{max}$/e after light propagates another 0.49 $\mu$m along the X axis. We cannot observe a S-shaped structure between the SP and EP in the formed photonic hook. As shown in Figure~\ref{fig:bjdx}(b), the focal length increases when the background RI is increased to n=1.52. The patchy particle forms a bended jet-like optical field with a small degree of curvature.  The beam has the strongest intensity (I$_{max}$=5.39) at X=1.40 $\mu$m, Y=-0.06 $\mu$m and reaches its EP at X=2.75 $\mu$m, Y=0.03 $\mu$m. We found that the patchy microcylinder generates S-shaped PHs with a better quality in water [Figure~\ref{fig:pnjvsph}(b)]. As water is a commonly used biocompatible material, studying PHs in water is important for the practical applications of such curved beams. We will focus on the generation of S-PHs in water in the following studies.

The influence of particle diameter on the S-PHs is also investigated in this study. As shown in Figures~\ref{fig:bjdx}(c)-(f), the I$_{max}$ position moves further away from the cylinder when the particle diameter increases from 1 $\mu$m to 5 $\mu$m. Larger particles also have a higher I$_{max}$ value because they can focus more energy into the hot spot. S-PHs only can be generated from the 4 $\mu$m-diameter patchy cylinder with a small curvature of $\beta_1$=13.8$^\circ$, $\beta_2$=14.6$^\circ$, which is much less curved than the S-PHs from the 2 $\mu$m-diameter microcylinder [Figure~\ref{fig:pnjvsph}(b)].

\begin{figure}[htb]
\centering
\includegraphics[width=\textwidth]{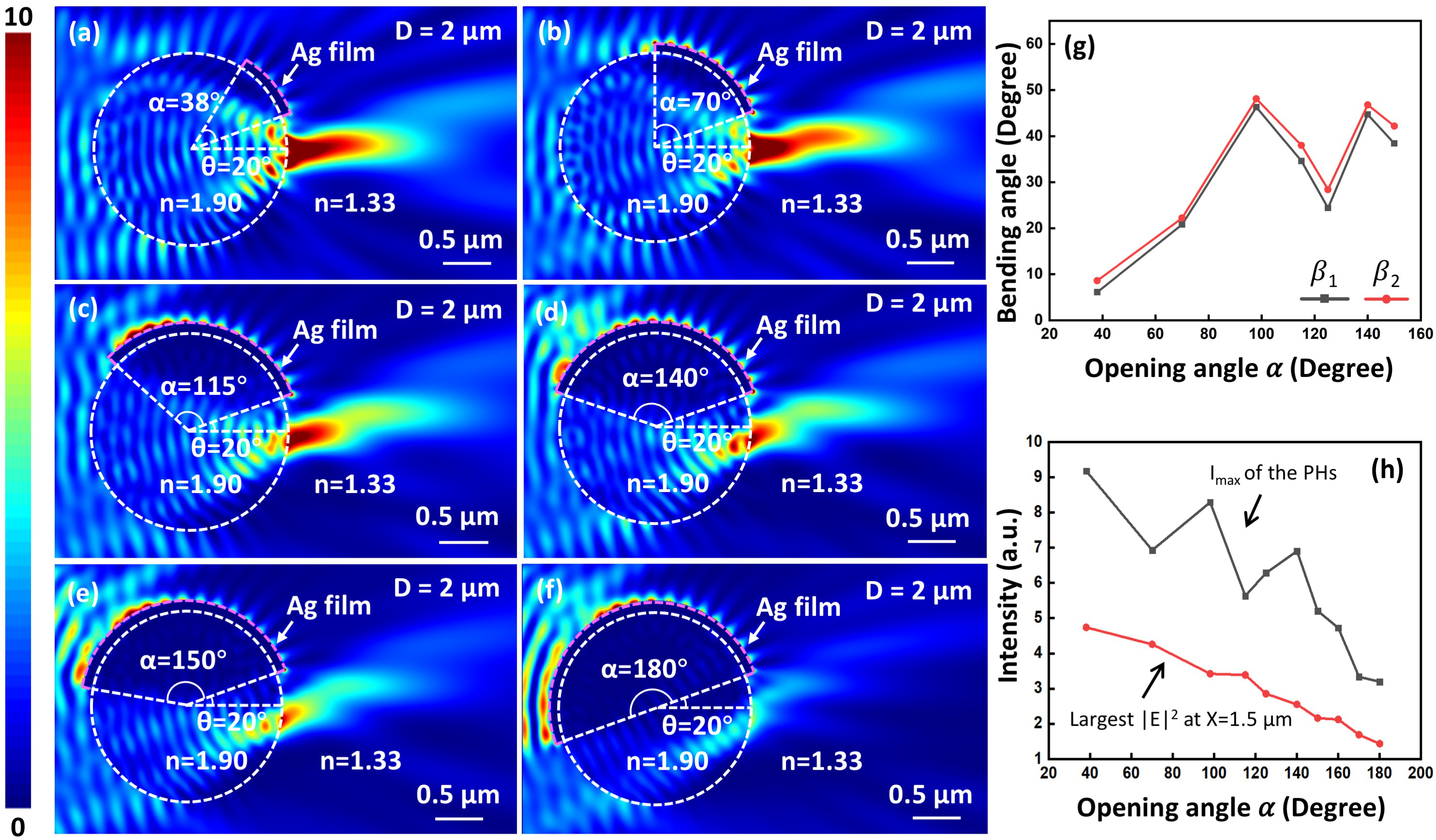}
\caption{(a)-(f) Optical fields formed by 2 $\mu$m-diameter patchy particles in water. The right edge of the Ag films are kept constant at $\theta$=20$^\circ$, while the opening angle ($\alpha$) of the Ag films increases from $\alpha$=38$^\circ$ to $\alpha$=180$^\circ$; (g) Bending angles of the PH as a function of the opening angle of Ag films; (h) I$_{max}$ of the PH (black line) and the maximum ${\mid}E{\mid}^2$ of the PH at X=1.5 $\mu$m (red line) as a function of the opening angle $\alpha$.}
\label{fig:dpy}
\end{figure}

As shown in Figure~\ref{fig:dpy}, changing the opening angle of Ag films can effectively affect the characteristics of the obtained PHs. Figures~\ref{fig:dpy}(a)-(f) show the optical fields generated by 2 $\mu$m-diameter patchy microcylinders in water. The opening angle of Ag films changes from $\alpha$=38$^\circ$ to $\alpha$=180$^\circ$ and the right edge of the Ag films are kept constant at $\theta$=20$^\circ$. As shown in Figures~\ref{fig:dpy}(h),(g), the PH has the strongest intensity enhancement ($I_{max}$=9.18) with a relatively small curvature ($\beta_1$=6.1$^\circ$, $\beta_2$=8.6$^\circ$) at $\alpha$=38$^\circ$ [Figure~\ref{fig:dpy}(a)]. The curvature of PHs becomes larger when the $\alpha$ increases to 98$^\circ$, and its curvature drops quickly to $\beta_1$=24.4$^\circ$ and $\beta_2$=28.4$^\circ$ at $\alpha$=125$^\circ$. The bending angles finally reach to $\beta_1$=38.4$^\circ$ and $\beta_2$=42.2$^\circ$ at $\alpha$=150$^\circ$. S-PHs cannot be found in the patchy particle with an opening angle larger than 150$^\circ$, because most of the energy concentrated in the hot spot will disperse quickly into the free space. The maxium curvature obtained in this configuration is $\beta_1$=46.3$^\circ$, $\beta_2$=48.1$^\circ$ at $\alpha$=98$^\circ$ [Figure~\ref{fig:dpy}(g)].

As shown in the black line in Figure~\ref{fig:dpy}(h), the particle with a larger opening angle tend to have a smaller I$_{max}$. However, we can see some turbulence in the plot between $\alpha$=70$^\circ$ and $\alpha$=150$^\circ$. The I$_{max}$ increases from 6.93 to 8.29 when the $\alpha$ increases from 70$^\circ$ to 98$^\circ$, and drops to 5.63 at $\alpha$=115$^\circ$. This turbulence also can be found when increasing $\alpha$ from 115$^\circ$ to 150$^\circ$. For all the opening angles, the I$_{max}$ points are always at the right edge of the particle. The largest I$_{max}$ obtained in this study is  9.18 at $\alpha$=38$^\circ$, and the minimum is 3.20 at $\alpha$=180$^\circ$. To avoid the turbulence caused by near-field effects, we also measured the maximum ${\mid}E{\mid}^2$ at X=1.5 $\mu$m, and thus a smoother curve can be obtained, as shown in the red line in Figure~\ref{fig:dpy}(h). We can see that the light intensity decreases approximately linearly from 4.74 to 1.43 as the opening angle $\alpha$ increases from 38$^\circ$ to 180$^\circ$. 

\begin{figure}[htb]
    \centering
    \includegraphics[width=\textwidth]{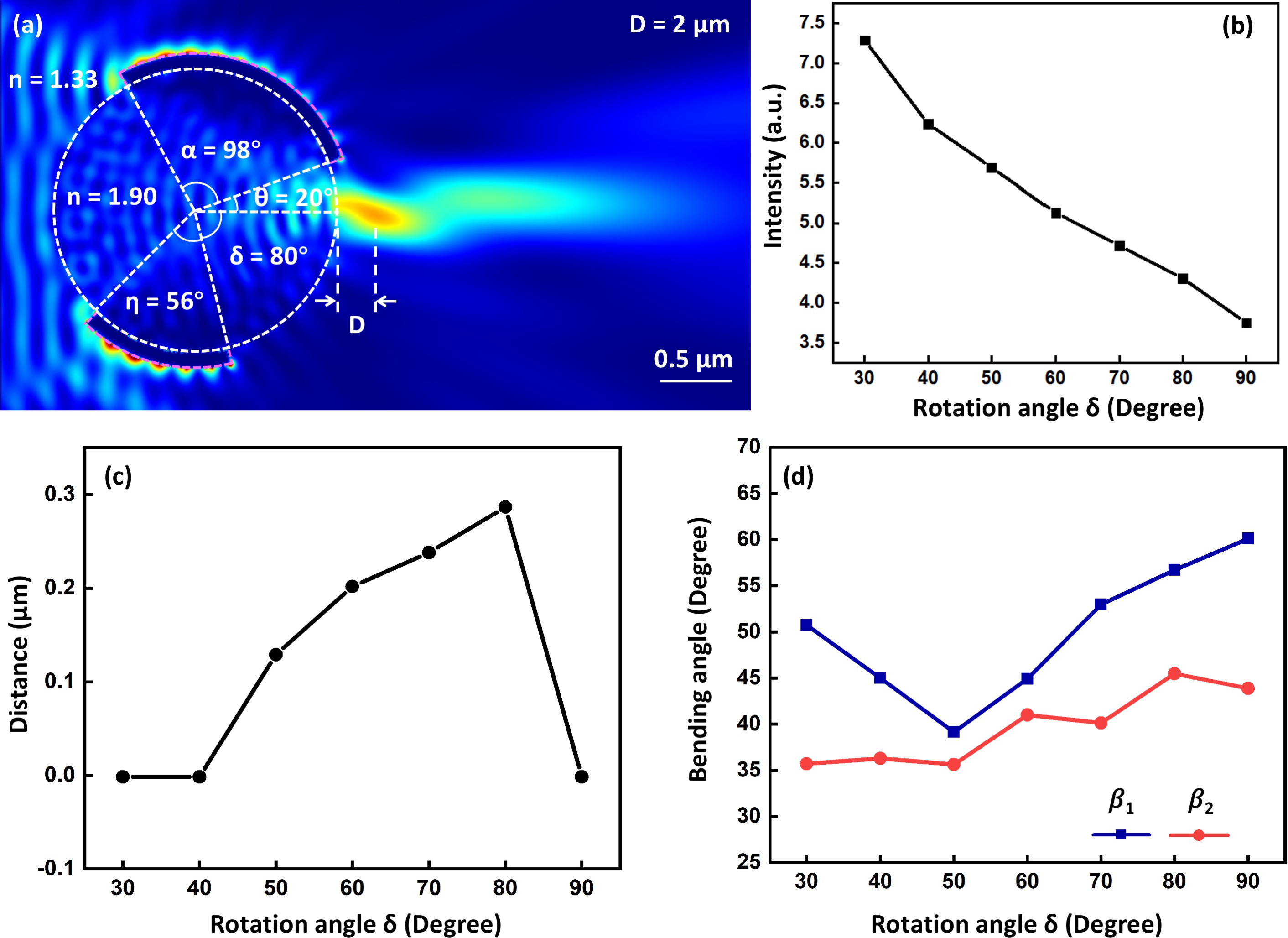}
    \caption{(a) Optical field of a patchy
    microcylinder with two patches of Ag films; (b)-(d) Influence of the rotation angle $\delta$ on the characteristics of the photonic hooks: (b) the value of I$_{max}$, (c) position of hot spot and (d) bending angles of photonic hooks obtained at different rotation angles.}
    \label{fig:twoAg} 
\end{figure}

Next, we studied the generation of S-PHs from microcylinders with two separate patches of Ag films.
As shown in Figure~\ref{fig:twoAg}(a), the two patches have an opening angle of $\alpha$ = 98$^\circ$ and $\eta$ = 56$^\circ$, respectively. The rotation angle of the patch on top is kept constant at $\theta$ = 20$^\circ$, and the rotation angle of the patch at bottom changes between $\delta$ = 30$^\circ$ to  $\delta$ = 90$^\circ$. 
The I$_{max}$ of the S-PHs decreases from 7.29 to 3.75 as the rotation angle $\delta$ increases from 30$^\circ$ to 90$^\circ$ [Figure~\ref{fig:twoAg}(b)].
The pupil of the microcylinder is smaller at larger $\delta$, leading to a smaller intensity enhancement factor.
The I$_{max}$ position of the S-PHs can be effectively adjusted by changing the $\delta$ angle.
The largest gap between the I$_{max}$ point and the microcylinder is $\sim$ 0.3 $\mu$m [Figure~\ref{fig:twoAg}(c)]. 
As shown in Figure~\ref{fig:twoAg}(d), the bending angle $\beta_1$ tends to decrease between $\delta$ = 30$^\circ$ ($\beta_1$ = 50.8$^\circ$) and $\delta$ = 50$^\circ$ ($\beta_1$ = 39.2$^\circ$), and then goes higher until $\delta$ = 90$^\circ$ ($\beta_1$ = 60.1$^\circ$). The bending angle $\beta_2$ increases slightly from $\beta_2$ = 35.7$^\circ$ to $\beta_2$ = 43.9$^\circ$ within this range.  The maximum curvature can be obtained at $\delta$ = 90$^\circ$ with $\beta_1$ = 60.1$^\circ$ and $\beta_2$ = 43.9$^\circ$.
Compared with the patchy particles covered by a single film, the particles with two patches can generate S-PHs with a much higher curvature and a tunable I$_{max}$ position.

\section{Conclusion}

We proposed a method for the generation of S-PHs, a curved light field with S-shaped structures, which can be formed at the shadow side of a patchy particle due to the asymmetric flow of energy when light wave propagates through the patchy particle.
The S-PHs studied in this work have a higher I$_{max}$ and smaller FWHM than conventional PNJs.
The characteristics of the S-PHs can be adjusted by changing the background refractive index, particle diameters and the coverage ratio of the patches.
Moreover, the curvature of the S-PHs can be significantly enhanced using microcylinders covered with two patches, and the position of the I$_{max}$ point can be adjusted by rotating the two patches. This new type of curved light may benefit various applications such as super-resolution imaging, micromachining, optical trapping, etc.

\printbibliography

\end{document}